\documentclass{ws-p8-50x6-00}

\begin{document}

\title{$J/\psi$ suppression in Pb--Pb collisions at CERN SPS}

\author{Francesco Prino for the NA50 collaboration}
\maketitle
\begin{center}
\begin{small}
M.C.~Abreu$^{6,a}$,
B.~Alessandro$^{10}$,
C.~Alexa$^{3}$,
R.~Arnaldi$^{10}$,
M.~Atayan$^{12}$,
C.~Baglin$^{1}$,
A.~Baldit$^{2}$,
M.~Bedjidian$^{11}$,
S.~Beol\`e$^{10}$,
V.~Boldea$^{3}$,
P.~Bordalo$^{6,b}$,
A.~Bussi\`ere$^{1}$,
L.~Capelli$^{11}$,
L.~Casagrande$^{6,c}$,
J.~Castor$^{2}$,
T.~Chambon$^{2}$,
B.~Chaurand$^{9}$,
I.~Chevrot$^{2}$,
B.~Cheynis$^{11}$,
E.~Chiavassa$^{10}$,
C.~Cical\`o$^{4}$,
T.~Claudino$^{6}$,
M.P.~Comets$^{8}$,
N.~Constans$^{9}$,
S.~Constantinescu$^{3}$,
N.~De Marco$^{10}$,
A.~De Falco$^{4}$,
G.~Dellacasa$^{10,d}$,
A.~Devaux$^{2}$,
S.~Dita$^{3}$,
O.~Drapier$^{11}$,
L.~Ducroux$^{11}$,
B.~Espagnon$^{2}$,
J.~Fargeix$^{2}$,
P.~Force$^{2}$,
M.~Gallio$^{10}$,
Y.K.~Gavrilov$^{7}$,
C.~Gerschel$^{8}$,
P.~Giubellino$^{10}$,
M.B.~Golubeva$^{7}$,
M.~Gonin$^{9}$,
A.A.~Grigorian$^{12}$,
J.Y.~Grossiord$^{11}$,
F.F.~Guber$^{7}$,
A.~Guichard$^{11}$,
H.~Gulkanyan$^{12}$,
R.~Hakobyan$^{12}$,
R.~Haroutunian$^{11}$,
M.~Idzik$^{10,e}$,
D.~Jouan$^{8}$,
T.L.~Karavitcheva$^{7}$,
L.~Kluberg$^{9}$,
A.B.~Kurepin$^{7}$,
Y.~Le~Bornec$^{8}$,
C.~Louren\c co$^{5}$,
P.~Macciotta$^{4}$,
M.~Mac~Cormick$^{8}$,
A.~Marzari-Chiesa$^{10}$,
M.~Masera$^{10}$,
A.~Masoni$^{4}$,
S.~Mehrabyan$^{12}$,
M.~Monteno$^{10}$,
A.~Musso$^{10}$,
P.~Petiau$^{9}$,
A.~Piccotti$^{10}$,
J.R.~Pizzi$^{11}$,
F.~Prino$^{10}$,
G.~Puddu$^{4}$,
C.~Quintans$^{6}$,
S.~Ramos$^{6,b}$,
L.~Ramello$^{10,d}$,
P.~Rato Mendes$^{6}$,
L.~Riccati$^{10}$,
A.~Romana$^{9}$,
I.~Ropotar$^{5}$,
P.~Saturnini$^{2}$,
E.~Scomparin$^{10}$
S.~Serci$^{4}$,
R.~Shahoyan$^{6,f}$,
S.~Silva$^{6}$,
M.~Sitta$^{10,d}$,
C.~Soave$^{10}$,
P.~Sonderegger$^{5,b}$,
X.~Tarrago$^{8}$,
N.S.~Topilskaya$^{7}$,
G.L.~Usai$^{4}$,
E.~Vercellin$^{10}$,
L.~Villatte$^{8}$,
N.~Willis$^{8}$.\\
\vspace{2mm}
$^{~1}$ LAPP, CNRS-IN2P3, Annecy-le-Vieux,  France.
$^{~2}$ LPC, Univ. Blaise Pascal and CNRS-IN2P3, Aubi\`ere, France.
$^{~3}$ IFA, Bucharest, Romania.
$^{~4}$ Universit\`a di Cagliari/INFN, Cagliari, Italy.
$^{~5}$ CERN, Geneva, Switzerland.
$^{~6}$ LIP, Lisbon, Portugal.
$^{~7}$ INR, Moscow, Russia.
$^{~8}$ IPN, Univ. de Paris-Sud and CNRS-IN2P3, Orsay, France.
$^{~9}$ LPNHE, Ecole Polytechnique and CNRS-IN2P3, Palaiseau, France.
$^{10}$ Universit\`a di Torino/INFN, Torino, Italy.
$^{11}$ IPN, Univ. Claude Bernard Lyon-I and CNRS-IN2P3, Villeurbanne,
France.
$^{12}$ YerPhI, Yerevan, Armenia.\\
\vspace{1mm}
a) also at UCEH, Universidade de Algarve, Faro, Portugal
b) also at IST, Universidade T\'ecnica de Lisboa, Lisbon, Portugal
c) now at CERN
d) Universit\'a del Piemonte Orientale, Alessandria and INFN-Torino,
Italy
e) now at Faculty of Physics and Nuclear Techniques, 
University of Mining and Metallurgy, Cracow, Poland
f) on leave of absence of YerPhI, Yerevan, Armenia\\
\end{small}
\end{center}

\abstracts{
The NA38 and NA50 experiments at the CERN SPS have measured charmonium 
production in different colliding systems with the aim of observing a phase 
transition from ordinary hadronic matter towards a state in which quarks and 
gluons are deconfined (Quark Gluon Plasma, QGP). 
In fact it was predicted that the $J/\psi$ yield should be suppressed 
in deconfined matter. 
The analysis of the data collected by the NA50 experiment with Pb--Pb 
collisions at 158 GeV/c per nucleon shows that the $J/\psi$ is anomalously 
suppressed in central collisions and the observed pattern can be considered as 
a strong indication for QGP production.}

\section{Introduction} 

In ordinary nuclear matter, quarks and gluons are confined inside nucleons.
Non-perturbative calculations of Quantum Chromo Dynamics predict that, 
when temperature exceeds a critical value T$_{c} \sim$ 150-180 MeV, nuclear
matter should undergo a phase transition into a state of matter in which 
quarks and gluons are no more confined into hadrons and behave as free 
particles. Such a state of matter is named Quark-Gluon Plasma (QGP).

Heavy-ion collisions are a powerful experimental tool to investigate 
nuclear matter under extreme conditions: the formation of the QGP is expected 
to occur in these collisions if the critical temperature and energy density 
required for the phase transition are reached.
Several probes have been proposed as signatures of the formation of a 
deconfined state of matter. 
In particular Matsui and Satz\cite{Satz86} predicted that the $J/\psi$ yield
would be suppressed in a deconfined medium due to the Debye screening of the 
attractive colour force which binds the {\em c} and $\bar{c}$ quarks together.
$J/\psi$ suppression is a particularly interesting signature of QGP formation
because it probes the state of matter in the earliest stages of the collision, 
since $c\bar{c}$ pairs can only be produced at that time. 
Moreover the $J/\psi$ is a tightly bound state that can not be easily broken 
by interactions with the hadronic medium and therefore it carries its original 
message through the different stages of the reacting medium.

At the CERN SPS, the NA38 and NA50 experiments have studied $J/\psi$ production
using the $\mu^+\mu^-$ decay channel with incident proton, oxygen, sulphur and 
lead ions on several targets. 

\section{Experimental setup and data taking conditions} 
 
The $J/\psi$ is detected via its $\mu^+\mu^-$ decay by a dimuon spectrometer
which consists of an air-gap toroidal magnet equipped with two sets of
multiwire proportional chambers (4 MWPC upstream the magnet and 4 downstream) 
for muon tracking purposes and with 6 scintillator hodoscopes to provide 
the dimuon trigger. 
The covered rapidity window is $2.8\leq y_{lab}\leq 4.0$.
 
The centrality of the collision is estimated event by event using three 
different detectors. 
The electromagnetic calorimeter 
measures the neutral transverse energy ($E_T$) produced in the interaction 
in the pseudorapidity window $1.1\leq \eta_{lab} \leq 2.3$. 
The zero degree calorimeter measures the energy $E_{ZDC}$, essentially carried 
by the projectile spectator (non-interacting) nucleons. 
It covers the pseudorapidity interval $\eta_{lab}\geq 6.3$. 
Finally the silicon microstrip detector measures the multiplicity and the
angular distribution of charged particles in the acceptance window 
$1.5 \leq \eta_{lab} \leq 3.5$.  

A quartz beam hodoscope (BH), placed about 33~m upstream the target is also used to 
count the incident lead ions and to reject beam pile-up.  
The NA50 target is a segmented active Pb target: there can be up to 7 
subtargets separated by few centimeters of air.
Two quartz blades located off the beam axis on the left and right side of each 
subtarget identify where the interaction has occurred.  
For a more detailed description of the detectors see Ref. 4. 


\section{Charmonium production in p--A and light ion interactions} 

NA38 and NA51 experiments collected an extensive set of measurements of 
$J/\psi$ production using p, O and S beams on several targets.
Since charmonium production is a hard process and therefore it scales as 
A$\times$B (where A and B are the projectile and the target mass number 
respectively), it is possible to define a cross-section per nucleon-nucleon 
collision as B$_{\mu \mu}\sigma_{J/\psi}/(A \times B)$, where B$_{\mu \mu}$
is the branching ratio of $J/\psi$ into two muons, in order to compare
charmonium production in different colliding systems.

\begin{figure}[htb]
\begin{center}
\resizebox{0.6\textwidth}{!}          
{\includegraphics*[bb= 0 0 545 554]{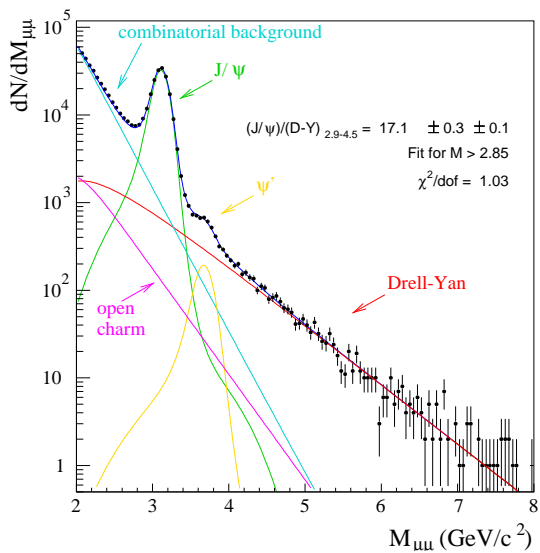}}          
\caption[uno]{Opposite sign muon pair invariant mass spectrum for Pb--Pb
collisions.} \label{fig:mass}
\end{center}
\end{figure}

The standard analysis method has been described in detail in Ref. 4. 
The number of J/$\psi$ events is determined by fitting the invariant mass 
spectrum of opposite sign muon pairs above 2.9 GeV/c$^2$.
The fit is performed including 5 contributions: $J/\psi$, $\psi'$, 
Drell-Yan, open charm (i.e. semileptonic decays of D and $\bar{D}$ mesons)
and combinatorial background from $\pi$ and K decays. 
An example of invariant mass spectrum is presented in fig.~\ref{fig:mass}
 
\begin{figure}[htb]
\begin{center}
\resizebox{0.6\textwidth}{!}          
{\includegraphics*[bb= 0 0 552 559]{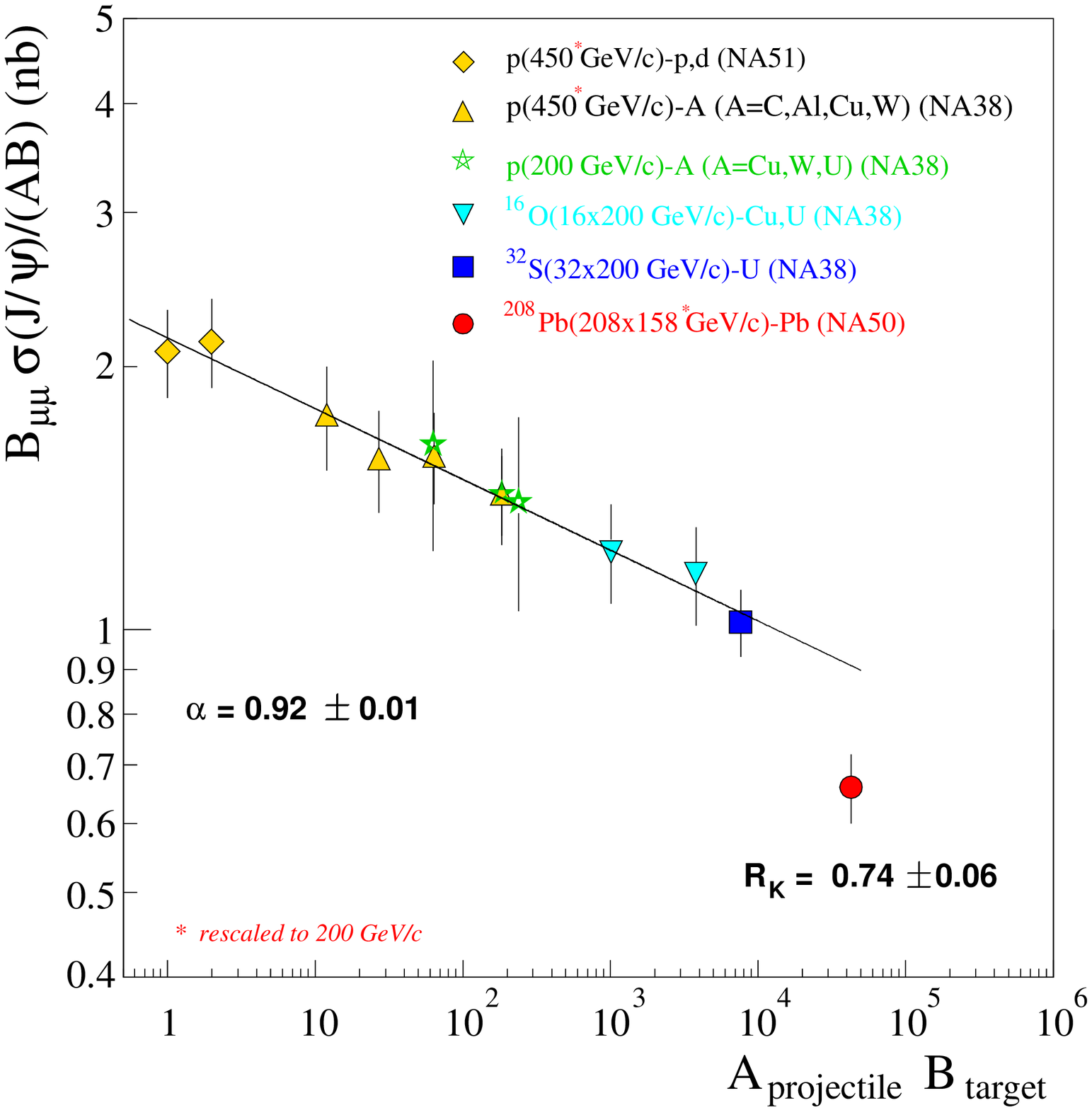}}          
\caption[uno]{$J/\psi$ cross section as a function of A$\times$B from p--p to
Pb-Pb interactions.} \label{fig:psi95}
\end{center}
\end{figure}

In figure~\ref{fig:psi95} the $J/\psi$ cross-section per nucleon-nucleon 
collision as a function of A$\times$B is represented: it can be seen that 
all the data from p-p up to central S-U collisions show a continuous 
and monotonic $J/\psi$ suppression pattern from the lighter to the heavier 
interacting nuclei\cite{NA51,NA38}.
The systematics of $J/\psi$ production from p-p to S-U can be parametrized by 
the simple law $\sigma_{AB}= \sigma_{0}(AB)^{\alpha}$ 
with $\alpha = 0.918\pm0.015$\cite{NA38}.
Within the framework of the Glauber model, using a simple first order 
exponential fit, these data lead to a $J/\psi$ absorption cross-section of 
$5.9\pm0.6$ mb (equivalent to $6.4\pm0.8$ mb for a full Glauber 
calculation)\cite{NA38}.
This result can be understood in terms of ordinary nuclear absorption of 
a preresonant $c\bar{c}$ state meant to become later on, if not destroyed, 
the fully formed $J/\psi$. 
This normal suppression sets the baseline with respect to which we can compare 
the pattern of $J/\psi$ production in Pb-Pb interactions.

\section{Anomalous J/$\psi$ suppression in Pb--Pb collisions} 

In fig.~\ref{fig:psi95} it can be seen that the $J/\psi$ cross-section 
measured in Pb--Pb collisions lies $\sim$ 5 standard deviations below the value
expected from the fit to the data from p--p to S--U and this result indicates 
that there is a new suppression mechanism at work.

In order to perform a study of charmonium production as a function of the 
collision centrality, it is necessary to find a replacement for A$\times$B 
in the definition of the cross section per nucleon-nucleon collision.
Since the Drell-Yan (DY) process provides muons in the same mass range as the 
$J/\psi$ and its cross section is expected to be proportional to the number 
of elementary nucleon-nucleon collisions without any sizeable nuclear effect, 
the ratio $B_{\mu \mu}\sigma_{J/\psi}/\sigma_{DY}$ can be used to study the 
centrality dependence of the $J/\psi$ suppression\cite{Abreu972} with the
advantage that in the ratio $\sigma_{J/\psi}/\sigma_{DY}$ systematic errors 
related to detector inefficiencies and flux uncertainties cancel. 
The transverse energy spectrum is then divided into bins: in each of these 
bins a fit to the invariant mass spectrum of the muon pairs is performed and 
the $J/\psi$ over DY ratio is calculated\cite{Abreu99}.

The disadvantage of this analysis method is that there are statistical 
fluctuations in the $\sigma_{J/\psi}/\sigma_{DY}$ ratio, essentially due to 
the small statistics of the DY sample. 
In order to overcome this problem, a new independent analysis method has been 
developed\cite{Abreu00} in which the sample of DY events is replaced by the 
huge sample of minimum bias (MB) events, collected with a beam trigger which 
fires every time that a non zero energy deposit is detected by the Zero Degree
Calorimeter (even for the most central collisions some energy, from produced 
particles, falls into the ZDC acceptance). 

In this anaysis method the number of $J/\psi$ events is obtained without any 
fitting procedure, simply by counting the number of opposite sign muon 
pairs in the mass range from 2.9 to 3.3 $\:GeV/c^2$, after combinatorial 
background subtraction.
In order to compare the results of this analysis with the ones obtained with 
the standard method, the minimum bias reference has to be converted into the 
Drell-Yan reference.
This is done multiplying the measured E$_{T}$ spectrum of the MB events 
by the ratio of the theoretical shapes of the E$_{T}$ spectra of
DY and MB, evaluated in the frame of the Glauber model.

Because of the bigger size of the minimum bias sample, the new analysis is no 
more affected by statistical fluctuations and a finer $E_T$ binning is possible. 
It is also free from most inefficiencies as it is still computed from a ratio 
of experimental numbers. 

\begin{figure}[htb]
\begin{center}
\resizebox{0.6\textwidth}{!} %
{\includegraphics*[bb= 0 0 552 601]{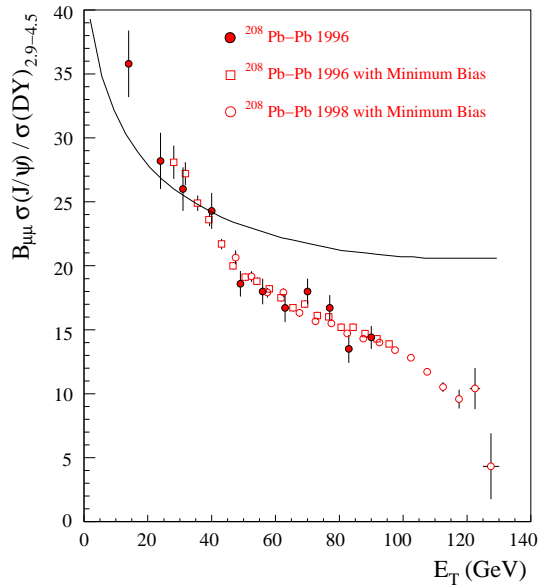}}          
\caption[uno]{$B_{\mu \mu}\sigma_{J/\psi}/\sigma_{DY}$ ratio as a function of $E_T$,
obtained with the standard and minimum bias analysis of the 1996 and 1998 data
samples.}\label{fig:psi98}
\end{center}
\end{figure}

The $J/\psi$ over DY ratios obtained with the two independent analysis show a
good agreement, as it can be seen in fig.~\ref{fig:psi98} where the 
complete $J/\psi$ suppression pattern obtained with the standard 
(closed circles) and the minimum bias (open points) analysis for the 1996 and 
1998 data samples is plotted as a function of the transverse energy E$_{T}$.

The solid line visible in the figure represents the ``ordinary'' nuclear 
absorption of the $c\bar{c}$ pair, which accounts for the $J/\psi$ yield 
measured in lighter colliding systems (from p-p to S-U). 
It can be observed that for the most peripheral Pb--Pb collisions the $J/\psi$
suppression can be accounted for by ordinary nuclear absorption while 
a clear departure from this trend can be seen at $E_T \approx$ 40 GeV, 
which approximately corresponds to an impact parameter of 8 fm.
It is also evident that there is no saturation of $J/\psi$ suppression for the 
most central Pb--Pb collisions: a second drop in the $J/\psi$ suppression 
pattern can be observed at $E_{T} \approx$ 90 GeV. 


\section{Conclusions}

The combined results of the NA38 and NA50 experiments clearly indicate a 
step-wise $J/\psi$ suppression pattern with no saturation for the most 
central collisions. 

The experimentally observed suppression is much steeper (when the experimental
E$_{T}$ resolution is taken into account) than the results of the predictions 
of all presently available models of  $J/\psi$ suppression based on the 
absorption of the meson by interactions with the surrounding hadronic matter. 

On the contrary the observed suppression pattern can be naturally understood 
in a deconfinement scenario.
Since $\approx$ 30--40\% of the $J/\psi$'s come from radiative decay of
$\chi_c$ mesons, the first anomalous step in charmonium suppression can be 
understood as due to the melting in deconfined matter of the $\chi_c$, 
which is less tightly bound than the $J/\psi$.
In this scenario, the second drop, observed for more central collisions, 
would be due to the dissolving of the more tightly bound $J/\psi$ meson, which
requires energy densities grater than the ones necessary to melt the $\chi_c$.

Hence it can be concluded that the $J/\psi$ suppression pattern observed in the
NA50 data provides significant evidence for deconfinement of quarks and gluons 
in Pb--Pb collisions.

\end{document}